# Criterions of Phase Transitions in Dispersed Multiphase Systems Based on an Extended Lattice Model

Yiran Li, Yunfan Huang, Xukang Lu, Moran Wang[†]

*Department of Engineering Mechanics, Tsinghua University, Beijing 100084, China*

## Abstract

Agglomeration, adsorption, and extraction in dispersed multiphase systems are ubiquitously encountered in biological systems, energy industry, and medical science. In this work, a novel lattice model is extended to the three-component complex systems and criterions based on the determinant $\Delta_i = F_i - K_{c,i}$ are accordingly proposed to predict the aforementioned behaviors based on a Helmholtz free energy formulation. Here, three characteristic factors $F_i$'s are introduced to describe the internal energy effect, i.e., $F_1 = A^{11} + A^{22} - 2A^{12}$ (agglomeration), $F_2 = A^{22} + A^{13} - A^{12} - A^{23}$ (adsorption) and $F_3 = A^{22} + 2A^{13} - A^{33} - 2A^{12}$ (extraction), where $A_{ij}$ denotes the conservative potential coefficient between liquid particles in phase *i* and *j*, while the entropy factors $K_{c,i}$'s in the determinants depend on the local structure of the liquid. To verify the theoretical criterions, many-body dissipative particle dynamics (mDPD) are employed to simulate the phase transition phenomena in various many-body dissipative systems, including agglomeration in a binary system, adsorption on solid surfaces from the liquid phase, and extraction-adsorption around an immiscible liquid-liquid interface. The simulation results show the notable dependence of the dispersed phase distribution transition on the quasi-order parameters $\Delta_i$'s, which indicates the great potential of our model to analyzing various dispersed multiphase systems. The criterions are expected to be extended to study the structure effect induced by the temperature and dispersed particle shape on the phase transition phenomena in those complex systems.

---

[†] Corresponding author. E-mail: mrwang@tsinghua.edu.cn

# I Introduction

Dispersed multiphase systems are mixtures consisting of immiscible phases with at least one component existing as discrete phase [1,2]. These mixtures are ubiquitous in biological systems, energy industry, and medical science. The distributions of the dispersed components in the mixtures such as blood cells and platelets, nanoparticles in nanofluids, nano-micelles in nano-emulsions have substantial impacts on human's cardiovascular system [3], oil recovery efficiency [4], and dermatological treatment [5]. Among various distributions of dispersed particles, agglomeration, adsorption, and extraction between different phases have drawn an increased interest among researchers. For example, it have been revealed that the shape, surface property, and charge of nanoparticles, as well as the chemical structure and polarities of solvent molecules are influential in the interaction between particles and continuous phases, hence the adsorption rate and structure at liquid-liquid interfaces [6-8] and solid surfaces [9]. In addition to the adsorption processes, the particle extraction through liquid-liquid interface, which is crucial in nanotechnology such as quantum dots and nanoparticles, has been intensively investigated [10-12]. Extractors with different sizes, structures and functional groups were synthesized to adjust the interaction at the solid-liquid interface, inducing the phase transfer of nanoparticles [10,12,13]. Although plenty of efforts have been devoted to the study of three typical distributions in multiphase systems, most of the existing theories and models are limited to specific systems [7,9,14-17], resulting in the limitation of lacking universal applicability. A general theory of dispersed multiphase system including analysis for all the three distributions is thus needed.

In essence, a dispersed multiphase system is a thermodynamic system consisting of interacting particles of different types, based on which an adsorption theory that is applicable to multiple binary mixtures including water-sulfuric acid, hydrogen-palladium, argon-graphite, etc., was proposed by Hill in 1949 [18], enabling the calculation of thermodynamic functions of the systems from experimental data. Despite its ability to quantitatively describe the thermodynamic properties of the system, this theory does not provide enough insight into the relationship between microscopic interactions and macroscopic distributions of particles, which is critical for the design of novel nanomaterials [4,8,10]. Theoretically, the equilibrium distribution of the dispersed phases can be determined by locating the minimum free energy point in the phase space of the whole system when all the particle interactions are known. Nevertheless, due to the unstructured nature of the mixture, it is impracticable to obtain the free energy of the system analytically.

Certain approximation methods have been adopted by previous researchers to construct simplified models for theoretical analysis of complex systems, among which the lattice model is a competitive candidate, such as Ising model describing the phase transition of physical systems and lattice gas cellular automata (LGCA) describing the emergent behaviour of fluid transport [19-21]. An idealized lattice model for binary mixtures was first constructed by Meyer, assuming a lattice structure in the liquid with lattice constants independent of composition [22], from which the correct entropy of mixing for two ideal liquids of different species with same number densities was derived [23]. Further modifications of Meyer's model, including regular solution theory and Flory-Huggins theory, were proposed and succeeded in explaining phase separation phenomena in nonideal solutions and polymer mixtures [22-24]. However, to the best of our knowledge, all the available lattice models are confined to binary systems and thus not applicable to investigate interfacial adsorption and phase transfer process of particles, which indeed constructs a ternary mixture. In the present work, we extend the lattice model to describe multicomponent systems in thermodynamic equilibrium and propose a method for approximating the free energy of the real system, whereby semi-quantitative criteria of phase change including the three typical distribution above expressed in interparticle potential parameters are derived in a general form. Further, many-body dissipative particle dynamics (mDPD) is employed to model binary and ternary mixtures for the verification of our approximation method.

The paper is organized as follows. In Sec. II, our extended lattice model is presented and the three relationships between interparticle interactions and particle distributions, i.e., agglomeration, adsorption, and extraction, are derived under different assumptions. Sec. III combines our lattice model with mDPD system and presents three criteria of the phase transition for the three distributions. Sec. IV shows the simulation settings and results to verify our theoretical analysis, with discussions about our model, following by the conclusion in Sec. V.

## II  Theoretical Formulation: An Extended Lattice Model

In this section, a novel lattice model with a global representation of the system free energy is presented as an extension of Meyer's model. Further, under the assumption of locality, the general form of Helmholtz free energy formulation is applied to obtain the criterions of three typical phase transition phenomena, i.e., agglomeration, adsorption, and extraction. The influence of the assumptions on the locality criterion is then carefully discussed, followed by a simplified version of the criterions, which will be verified in the next sections.

### II.A    Global Helmholtz free energy formulation

Consider an equilibrium system composed of $M$ types of particles similar in size with the particle number, volume, and absolute temperature being constant. Conforming to Meyer's model, each particle is assumed to occupy one lattice site without overlapping on a hypothetical simple cubic lattice. While being a relatively independent quantity in Meyer's model, here the lattice constant $a$ is uniquely determined by the average number density $n$ of the real system as a geometrical correspondence, satisfying

$$\frac{1}{a^3} = n, \qquad (1)$$

which is analogous to the treatment of the cell model in colloid science[25,26]. Instead of restricting the interparticle interactions to neighboring particles as Meyer did, we consider the interactions of all particle pairs in our model. Since the lattice structure of the system is locally assumed, the distance between particles is discretely distributed and thus can be ordered. In the following, we denote the $\alpha$-th closest distance between lattice sites by $r_\alpha$, and the corresponding number of neighbors at that distance for each site by $c_\alpha$. If a particle pair consists of a type $i$ particle and a type $j$ particle with spacing $r_\alpha$, we call it an "$i$-$j$ type $\alpha$-neighbor", the interactional energy and total number of which are denoted by $u_\alpha^{ij}$ and $N_\alpha^{ij}$, respectively.

The total interaction energy $U^{\text{pot}}$ can then be written as

$$U^{\text{pot}} = \sum_{\alpha=1}^{\infty} \sum_{i=1}^{M} \sum_{j=1}^{i} N_{\alpha}^{ij} \cdot u_{\alpha}^{ij}, \tag{2}$$

which indicates that $U^{\text{pot}}$ is determined by both the number $N_{\alpha}^{ij}$ and the energy $u_{\alpha}^{ij}$ of each type of pair. In contrast, the entropy of the system $S \equiv S(\{N_{\alpha}^{ij}\})$, as a geometric quantity, is only dependent on the set $\{N_{\alpha}^{ij}\}$. Defining two different configurations A and B by giving different sets, $N_{\alpha}^{ij}$ and $N_{\alpha}^{ij'}$, respectively, we obtain the difference between their total interaction energies and entropies, namely

$$\Delta U^{\text{pot}} = U^{\text{pot}} - U^{\text{pot}'} = \sum_{\alpha=1}^{\infty} \sum_{i=1}^{M} \sum_{j=1}^{l} \left( N_{\alpha}^{ij} - N_{\alpha}^{ij'} \right) \cdot u_{\alpha}^{ij}, \tag{3}$$

$$\Delta S = S - S' = S(\{N_{\alpha}^{ij}\}) - S(\{N_{\alpha}^{ij'}\}). \tag{4}$$

Assuming that the internal and kinetic energy of particles are independent of interparticle interactions, as in Meyer's model, we obtain the difference in Helmholtz free energy between the configuration A and B

$$\Delta F = \Delta U^{\text{total}} - T\Delta S = \Delta U^{\text{pot}} - T\Delta S$$
$$= \sum_{\alpha=1}^{\infty} \sum_{i=1}^{M} \sum_{j=1}^{i} \left( N_{\alpha}^{ij} - N_{\alpha}^{ij'} \right) \cdot u_{\alpha}^{ij} - T\left( S(\{N_{\alpha}^{ij}\}) - S(\{N_{\alpha}^{ij'}\}) \right). \tag{5}$$

In the canonical ensemble, the preferrable configuration of the system is determined by the sign of $\Delta F$; if $\Delta F < 0$, configuration A is preferred, and vice versa.

As shown in Eq. (5), the system configuration is resulted from the competition between the energy effect and the entropy effect, determined by the interaction potential set $\{u_{\alpha}^{ij}\}$ and the configurations $\{N_{\alpha}^{ij}\}$. Although the exact form of $\Delta S$ is difficult to obtain, it will be presented in subsequent sections that significant information can still be provided by Eq. (5) to give rather simple determinants consisting of potential parameters to predict and control the particle distribution. In this contribution, we focus on the energy effect and aim to correlate the complex phase transition phenomena in multicomponent system with explicit criterions.

## II.B  General criterion of phase transition under locality assumption

To move on, we need to adopt some assumptions on the locality and interparticle potential. In this subsection, we neglect all the interparticle interactions between $\alpha$-neighbors ($\alpha > \alpha_0 = 1$) as in the Meyer's model for simplicity (which we call a "$\alpha_0$-locality" assumption and $\alpha_0$ is named the "locality index"), and this restriction will be relaxed in the next subsection.

Under the 1-locality assumption, Eq. (5) is simplified into the following form

$$\Delta F = \Delta U^{total} - T\Delta S = \Delta U^{pot} - T\Delta S$$
$$= \sum_{i=1}^{M}\sum_{j=1}^{i}\left(N_1^{ij} - N_1^{ij'}\right) \cdot u_1^{ij} - T\left(S\left(\{N_1^{ij}\}\right) - S\left(\{N_1^{ij'}\}\right)\right), \quad (6)$$

which is then employed to investigate the criterions of phase transition for three types of distributions: agglomeration, adsorption and extraction. Following the statistical treatment in [23], the interaction potential term is reorganized as the function of independent variables under the constraints of certain geometric relations. The Helmholtz free energy differences for binary and ternary systems can be rewritten as

$$\Delta F = -\left(u_1^{11} + u_1^{22} - 2u_1^{12}\right)\frac{\Delta N_1^{12}}{2} - T\left(S\left(\{N_1^{ij}\}\right) - S\left(\{N_1^{ij'}\}\right)\right), \quad (7)$$

$$\Delta F = \sum_{\substack{m,n=1\\m\neq n}}^{3}\left[-\left(u_1^{mm} + u_1^{nn} - 2u_1^{mn}\right)\frac{\Delta N_1^{mn}}{2}\right] - T\left(S\left(\{N_1^{ij}\}\right) - S\left(\{N_1^{ij'}\}\right)\right), \quad (8)$$

respectively, where $\Delta N_1^{mn} = N_1^{mn} - N_1^{mn'}$. It is noteworthy that in the latter two types of phase transition, we have adopted the additional assumption of dilute dispersed system and the existence of 1-1 type 1-neighbors is neglected, which reduces the degree of freedom in the analysis. The results are listed in Table 1, and the detailed calculations can be found in Appendix A.1-3.

As shown in Table 1, the determinants $\Delta_i^0 = F_i^0 - K_{c,i}^0$ are accordingly proposed to predict the aforementioned behaviors, behaving as the order parameters which we called the "quasi-order parameter". Here, three characteristic factors $F_i^0$'s are introduced to describe the internal energy effect represented by the linear combinations of conservative potential coefficient between liquid particles in phase $i$ and $j$, while the entropy factors $K_{c,i}^0$'s in the determinants depend on the local structure of the liquid. When the quasi-order parameters $\Delta_i^0 < 0$, the corresponding phenomena will happen remarkably. Note that the two-phase agglomeration factor for a binary mixtures here is consistent with the previous studies [22,23], while the latter two factors for adsorption and extraction are proposed for the first time. The mathematical form is quite physically intuitive, consistent with the operational thermodynamic considerations [27], but has a more solid statistical foundation with the flexibility for further extension. For example, the enabled parameter controlling and distribution prediction in the numerical modeling based on the estimation of the energy factor, as well as the theoretical prediction of the collective behavior in various dispersed systems with different local structures based on the explicit calculation of the entropy factors.

**Table** 1. Definition of terms in the quasi-order parameters $\Delta_i^0 = F_i^0 - K_{c,i}^0$ in the general criterion of phase transition phenomena under locality assumption.

| Number $i$ | Phenomena | Energy factor $F_i^0$ | Entropy factor $K_{c,i}^0$ |
|---|---|---|---|
| 1 | agglomeration | $u_1^{11} + u_1^{22} - 2u_1^{12}$ | $-\dfrac{T\Delta S}{\Delta N_1^{12}/2}$ |
| 2 | adsorption | $u_1^{22} + u_1^{13} - u_1^{12} - u_1^{23}$ | $-\dfrac{T\Delta S}{N^{(1)}}$ |
| 3 | extraction | $u_1^{22} + 2u_1^{13} - u_1^{33} - 2u_1^{12}$ | $-\dfrac{T\Delta S}{3N^{(1)}}$ |

## II.C    Relaxed locality assumption: conformal interparticle potential

Seldomly only the 1-neighbor interparticle interaction matters in the dispersed systems, taking the solvent particle in a nanoparticle suspension for example. To relax the 1-locality assumption, i.e., to increase the locality index $\alpha_0$, we need to impose certain assumption on the form of interparticle potential as a balance. Mean-field approximation is usually encountered in various fields of physics, especially in the condensed matter physics. Here, we introduce a conformal interparticle potential form, which could be recognized as a natural extension from the mean-field potential. In fact, it will be seen that once the conformal potential is supposed, the assumption of locality is no longer required to obtain a simple and elegant determinant in the criterion of phase transitions, i.e., $\alpha_0 = \infty$.

To be specific, the interparticle potential is assumed in the following form (which we called a "conformal" one)

$$u^{ij}(r) = A^{ij}w(r) + v(r), \qquad (9)$$

where $A^{ij}$ denotes the conservative potential coefficient between particles in phase $i$ and $j$, $w(r)$ and $v(r)$

are the identical weight functions for all particle pairs. The free energy differences between different configurations A and B of binary and ternary systems are then expressed as

$$\Delta F = \sum_{\alpha=1}^{\infty}\left(-\left(u_\alpha^{11}+u_\alpha^{22}-2u_\alpha^{12}\right)\frac{\Delta N_\alpha^{12}}{2}\right)-T\left(S\left(\{N_\alpha^{ij}\}\right)-S\left(\{N_\alpha^{ij'}\}\right)\right), \qquad (10)$$

$$\Delta F = \sum_{\alpha=1}^{\infty}\sum_{\substack{m,n=1\\m>n}}^{3}\left[-\left(u_\alpha^{mm}+u_\alpha^{nn}-2u_\alpha^{mn}\right)\frac{\Delta N_\alpha^{mn}}{2}\right]-T\left(S\left(\{N_\alpha^{ij}\}\right)-S\left(\{N_\alpha^{ij'}\}\right)\right), \qquad (11)$$

respectively. Substitute Eq. (12) into Eq. (10) and Eq. (11), we obtain

$$\Delta F = -(A^{11}+A^{22}-2A^{12})\sum_{\alpha=1}^{\infty}\left(w(r_\alpha)\frac{\Delta N_\alpha^{12}}{2}\right)-T\Delta S, \qquad (13)$$

$$\Delta F = \sum_{\substack{m,n=1\\m>n}}^{3}\left[-\left(A^{mm}+A^{nn}-2A^{mn}\right)\sum_{\alpha=1}^{\infty}w(r_\alpha)\frac{\Delta N_\alpha^{mn}}{2}\right]-T\Delta S, \qquad (14)$$

where $\Delta S = S\left(\{N_\alpha^{ij}\}\right)-S\left(\{N_\alpha^{ij'}\}\right)$.

Following a similar statistical treatment in the previous subsection, the results are obtained and listed below in Table 2. For the agglomeration phenomena, the calculation is quite straightforward in the same manner as the case under 1-locality assumption, while the detail calculations of the latter two could be found in Appendix A.4. The energy factors $F_i$'s here share the isomorphic forms with the ones in Table 1, which is the natural corollary by the conformal form of the interparticle potentials. As we will see in the next section, this formulation plays an important role in modeling and predicting the behaviors of the mDPD systems.

**Table** 2. Definition of terms in the quasi-order parameters $\Delta_i = F_i - K_{c,i}$ in the general criterion of phase transition phenomena under conformal potential assumption.

| Number $i$ | Phenomena | Energy factor $F_i$ | Entropy factor $K_{c,i}$ |
|---|---|---|---|
| 1 | agglomeration | $A^{11}+A^{22}-2A^{12}$ | $-\dfrac{T\Delta S}{\sum_{\alpha=1}^{\infty}\left(w(r_\alpha)\dfrac{\Delta N_\alpha^{12}}{2}\right)}$ |
| 2 | adsorption | $A^{22}+A^{13}-A^{12}-A^{23}$ | $-\dfrac{T\Delta S}{\sum_{\alpha=1}^{\infty}\left(w(r_\alpha)\Delta N_\alpha^{12}\right)}$ |
| 3 | extraction | $A^{22}+2A^{13}-A^{33}-2A^{12}$ | $-\dfrac{T\Delta S}{\sum_{\alpha=1}^{\infty}\left(w(r_\alpha)\dfrac{\Delta N_\alpha^{12}}{2}\right)}$ |

# III Application of Extended Lattice Model: mDPD systems

In this section, our lattice model and the above analysis are applied to many-body dissipative particle dynamics (mDPD) systems. A brief introduction to mDPD systems is first given, following by the translation of criterion of phase transition phenomena to the mDPD interaction coefficients. The criterions of agglomeration, adsorption and extraction derived here will be verified through mDPD simulations in Sec. IV.

## III.A   Short overview of the mDPD system

mDPD system consists of $N$ interacting particles, whose positions and velocities are denoted by $\boldsymbol{r}_i$ and $\boldsymbol{v}_i$ respectively, evolving under the control of Newtown's law of motion

$$\frac{d\boldsymbol{r}_i}{dt} = \boldsymbol{v}_i, \quad \frac{d\boldsymbol{v}_i}{dt} = \frac{\boldsymbol{F}_i}{m_i} \tag{15}$$

where $m_i$ is the mass of particle $i$. The total force acting on particle $i$, $\boldsymbol{F}_i$, is composed of external force $\boldsymbol{F}_i^{\text{ext}}$ and the sum of pairwise forces $\boldsymbol{F}_{ij}$ owing to the interaction between particle $i$ and particle $j$, namely

$$\boldsymbol{F}_i = \boldsymbol{F}_i^{\text{ext}} + \sum_{i \neq j} \boldsymbol{F}_{ij} \equiv \boldsymbol{F}_i^{\text{ext}} + \sum_{i \neq j} (\boldsymbol{F}_{ij}^{\text{C}} + \boldsymbol{F}_{ij}^{\text{D}} + \boldsymbol{F}_{ij}^{\text{R}}) \tag{16}$$

where the superscripts in the pairwise force terms denote the conservative force, dissipative force, and random force, respectively

$$\begin{aligned}
\boldsymbol{F}_{ij}^{\text{C}} &= A_{ij} w_c(r_{ij}) \hat{\boldsymbol{r}}_{ij} + B(\bar{\rho}_i + \bar{\rho}_j) w_d(r_{ij}) \hat{\boldsymbol{r}}_{ij}, \\
\boldsymbol{F}_{ij}^{\text{D}} &= -\gamma w_{\text{D}}(r_{ij})(\boldsymbol{v}_{ij} \cdot \hat{\boldsymbol{r}}_{ij}) \hat{\boldsymbol{r}}_{ij}, \\
\boldsymbol{F}_{ij}^{\text{R}} &= \sigma w_{\text{R}}(r_{ij}) \zeta_{ij} \hat{\boldsymbol{r}}_{ij}.
\end{aligned} \tag{17}$$

where $\hat{\boldsymbol{r}}_{ij} = \boldsymbol{r}_{ij}/r_{ij}$ and $\boldsymbol{v}_{ij} = \boldsymbol{v}_i - \boldsymbol{v}_j$. The weight functions in the expression of conservative force are determined according to previous mDPD studies [28-30], $w_c(r_{ij}) = (1 - r_{ij}/r_c) H(r_{ij} - r_c)$ and $w_d(r_{ij}) = (1 - r_{ij}/r_d) H(r_{ij} - r_d)$, which vanish when $r > r_c$ and $r > r_d$ ($H(x)$ is the Heaviside function), respectively, while $w_{\text{D}}$ and $w_{\text{R}}$ satisfying $w_{\text{D}} = w_c^2$ and $w_{\text{R}} = w_c$. Interparticle conservative potential parameter $A_{ij} < 0$ governs the magnitude of the attractive part which could differ for different type of particle pairs, while $B > 0$ controlling the amplitude of the repulsive part is necessary a constant for all particle pairs to ensure the mDPD system evolve as a Hamiltonian system [31]. The weighted density $\bar{\rho}_i$ is defined as

$$\bar{\rho}_i = \sum_{j \neq i} w_\rho(r_{ij}) \equiv \sum_{j \neq i} \frac{15}{2\pi r_d^3}\left(1-\frac{r_{ij}}{r_d}\right)^2 H(r_{ij}-r_d), \qquad (18)$$

which vanishes for $r > r_d$. $\gamma$ and $\sigma$ are the amplitudes of the dissipative and random forces. $\zeta_{ij}$ is a Gaussian random noise with $\langle \zeta_{ij}(t) \rangle \geq 0$ and $\langle \zeta_{ij}(t)\zeta_{kl}(t') \rangle \geq (\delta_{ik}\delta_{jl} + \delta_{il}\delta_{jk})\delta(t-t')$. In addition, the fluctuation-dissipation relation $\sigma^2 = 2\gamma k_B T$ is imposed on the random and dissipative coefficients to keep the equilibrium distribution of the mDPD system the same as that of a Hamiltonian system with conservative interactions $\boldsymbol{F}_i^C$ only [32], indicating the effectiveness of all the standard thermodynamic relations in mDPD systems, which is the prerequisite for the application of our lattice model.

### III.B  Extended lattice model of mDPD systems

In most mDPD models constructed in previous studies, the average number density $n$ ranges from 5 to 10. According to Eq. (1), the corresponding lattice constant $a$ is in the range of 0.45 to 0.6. Given that conventionally the cutoff distance $r_c$ is set to the unit of length in mDPD studies [32], namely, $r_c = 1 \doteq 2a$, $\{r_\alpha\}$ and $\{c_\alpha\}$ are thus determined as

$$r_1 = \frac{1}{2}r_c, \; r_2 = \frac{\sqrt{2}}{2}r_c, \; r_3 = \frac{\sqrt{3}}{2}r_c, \; c_1 = 6, \; c_2 = 12, \; c_3 = 8 \qquad (19)$$

where the 3-locality assumption has been adopted.

As mentioned before, the equilibrium distribution of the mDPD system is the same as that of a Hamiltonian system with conservative interactions $\boldsymbol{F}_i^C$ only. Hence, for mDPD particle pairs, the interactional energy could be obtained by integrating the conservative force against displacement

$$u^{ij}(r) = \int_r^\infty F_{ij}^C(\zeta)d\zeta = A_{ij}\int_r^\infty w_c(\zeta)d\zeta + B\int_r^\infty (\bar{\rho}_i + \bar{\rho}_j)w_d(\zeta)d\zeta, \qquad (20)$$

Since the number density is uniform and homogeneous $\bar{\rho}_i \equiv \bar{\rho}$ in the lattice model, $u_\alpha^{ij}$ can be further expressed as follows

$$u^{ij}(r) = \int_r^\infty F_{ij}^C(\zeta)d\zeta = A_{ij}\int_r^\infty w_c(\zeta)d\zeta + 2B\bar{\rho}\int_r^\infty w_d(\zeta)d\zeta. \qquad (21)$$

Eq. (21) shows that the mDPD interaction potential is in the conformal form (Eq. (9)) where

$$A^{ij} = A_{ij},$$
$$w(r) = \int_r^\infty w_c(\zeta)d\zeta, \qquad (22)$$
$$v(r) = 2B\bar{\rho}\int_r^\infty w_d(\zeta)d\zeta.$$

Hence, all the results presented in Sec. II.C is applicable to mDPD systems. In particular, the three mDPD coefficient combinations, which respectively control the phase transition of agglomeration, adsorption, and extraction are acquired

$$F_1 = A_{11} + A_{22} - 2A_{12},$$
$$F_2 = A_{22} + A_{13} - A_{12} - A_{23}, \qquad (23)$$
$$F_3 = A_{22} + 2A_{13} - A_{33} - 2A_{12}.$$

In addition, the transition points could be further expressed as:

$$K_{c,1} = -\frac{T\Delta S}{\sum_{\alpha=1}^\infty \left(w(r_\alpha)\frac{\Delta N_\alpha^{12}}{2}\right)} = -\frac{T\Delta S}{\frac{3}{8}\frac{\Delta N_1^{12}}{2} + \frac{-1+2\sqrt{2}}{4}\frac{\Delta N_2^{12}}{2} + \frac{-3+4\sqrt{3}}{8}\frac{\Delta N_3^{12}}{2}}$$

$$K_{c,2} = -\frac{T\Delta S}{\sum_{\alpha=1}^\infty \left(w(r_\alpha)\Delta N_\alpha^{12}\right)} == -\frac{T\Delta S}{\frac{3}{8}\Delta N_1^{12} + \frac{-1+2\sqrt{2}}{4}\Delta N_2^{12} + \frac{-3+4\sqrt{3}}{8}\Delta N_3^{12}} \qquad (24)$$

$$K_{c,3} = -\frac{T\Delta S}{\sum_{\alpha=1}^\infty \left(w(r_\alpha)\frac{\Delta N_\alpha^{12}}{2}\right)} == -\frac{T\Delta S}{\frac{3}{8}\frac{\Delta N_1^{12}}{2} + \frac{-1+2\sqrt{2}}{4}\frac{\Delta N_2^{12}}{2} + \frac{-3+4\sqrt{3}}{8}\frac{\Delta N_3^{12}}{2}}$$

Following the discussions in Appendix A, it can be inferred that $K_{c,1}$ and $K_{c,2}$ are all negative constants, while the sign of $K_{c,3}$ is determined by the ratio $N^{(2)}/N^{(3)}$. The analyses above provide important guidance for particle distribution control in mDPD systems. In other words, the occurrences of agglomeration, adsorption, and extraction should be determined by the combination of coefficients, which will be verified soon.

## IV MDPD Simulation: Results and Discussions

In this section, three different mDPD systems corresponding to the three extended models introduced in Sec. III are constructed. For each system, numerous simulations with different sets $\{A_{ij}\}$ are conducted to examine the relationships between particle distribution and interparticle interactions predicted by our model.

### IV.A Agglomeration

A binary mDPD system composed of 600 type 1 particles as dispersed phase and 7200 type-2 particles as continuous phase was constructed, with the simulation box of size $23.6 \times 2.0 \times 23.6$ in reduced units. The total particle density is about 7, satisfying the approximation condition $r_c \doteq 2a$. Periodic boundary conditions are applied in three dimensions. In the initial state, all type 1 particles were uniformly dispersed in phase 2 as shown in FIG. 1. Then, the system evolved to the equilibrium state.

The degree of agglomeration is quantified in a similar way as in a previous work [33], dividing the simulation box into $10 \times 1 \times 10$ blocks and calculating the relative standard deviation $D$ of the number density of type 1 particles in each block

$$D = \sqrt{\frac{\sum_{i=1}^{M}(n_i / n_{ave} - 1)^2}{M - 1}} \qquad (25)$$

where $M$ is the number of blocks, $n_i$ is the number density of type 1 particles in block $i$ and $n_{ave}$ is the average number density of type 1 particles. Then, simulations were conducted for 104 different models, where $F_1$ ranges from -30 to 30 with an interval of 5.

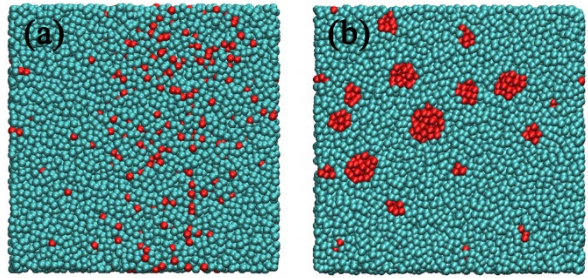

FIG. 1. The particle distribution: (a) without agglomeration; (b) with agglomeration.

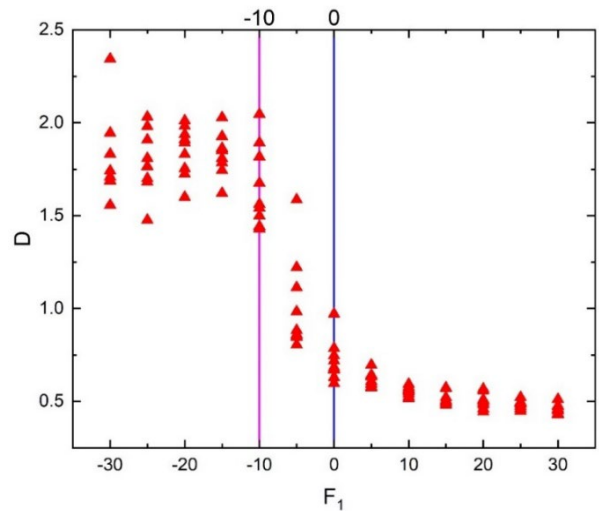

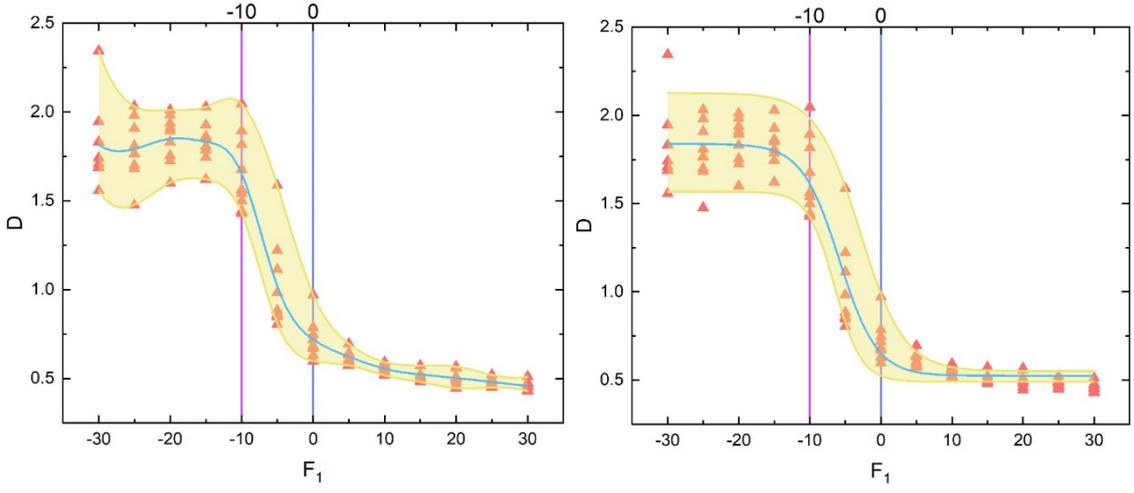

FIG. 2. The degree of agglomeration as a function of the agglomeration factor $F_1 = -2A_{12} + A_{11} + A_{22}$. It is seen that when $F_1 < -10$, the relative standard deviation $D$ of the number density of type 1 particles is scattered in the high-level range of 1.5 to 2.3, while when $F_1 > 0$, $D$ decreases and converges to about 0.5. For the transition zone where $-10 < F_1 < 0$, $D$ is distributed over a wide range from 0.6 to 2.0.

In FIG. 2, it is indicated that the transition of the particle distribution from significant agglomeration to uniform dispersion happens as $F_1$ increases, which is consistent with the predictions of our model. The transition point $K_c$ is in the interval [-10,0] and therefore negative, as estimated before. However, when $F_1$ is close to $K_c$, for example $F_1 = -5$, accurate predictions of the distribution become impossible, for which there are several possible reasons. The first is that in this situation, the difference in free energy $\Delta F$ between different distributions could be comparable to the thermal energy and no stable distribution can exist due to the random motions of particles. The inaccuracy could also be caused by the errors introduced by the simplifications in our model, and when $\Delta F$ is close to these errors, the prediction loses its accuracy.

## IV.B   Adsorption on Solid Surfaces

In our extended lattice model, the adsorption of particles at the liquid-liquid interface and on the solid surface is considered as the same process from the perspective of the Helmholtz free energy. However, in the real situation, adsorption of particles at the liquid-liquid interface may result in the phase transfer of particles and be coupled with the extraction process. Therefore, in this section, only adsorption on the solid surface is considered to investigate the adsorption process independently and adsorption at the liquid-liquid interfaced would be discussed together with extraction in the next section.

Here a ternary mDPD system composed of 600 type 1 particles as dispersed phase, 6400 type 2 particles as liquid phase and 7056 type 3 particles as solid phase, with the simulation box of size $33.0 \times 2.0 \times 33.0$ in reduced units, was constructed. Like the setup in Sec. VI.A, the approximation condition is fulfilled and the periodic boundary conditions are applied. After equilibrium, all solid particles were simultaneously attached to their respective sites by spring forces at a given moment and an additional linear repulsive force field was applied to all non-solid particles entering the solid region, as in the literature [34]. Initially, all type 1 particles were uniformly dispersed in phase 2 as shown in FIG. 3. Then, the system was allowed to evolve to the equilibrium state.

All the type 1 particles with a distance less than $r_c$ from the solid surface were regarded as being adsorbed. And the proportion of type 1 particles adsorbed on the solid surface $X$ was used to quantify the degree of adsorption

$$X = \frac{N^S}{N^{(1)}}, \qquad (26)$$

where $N^S$ is the number of type 1 particles adsorbed on the solid surface and $N^{(1)}$ is the total number of type 1 particles. The degrees of adsorption $X$ in 112 models with $F_2$ ranging from -100 to 30, satisfying $F_1 > 0$ to avoid agglomeration in phase 2, were calculated.

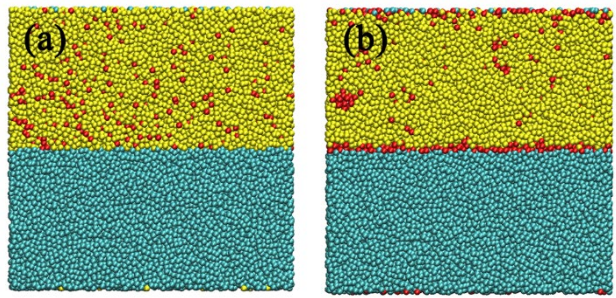

FIG. 3. The particle distribution: (a) without adsorption on solid surface; (b) with adsorption on solid surface.

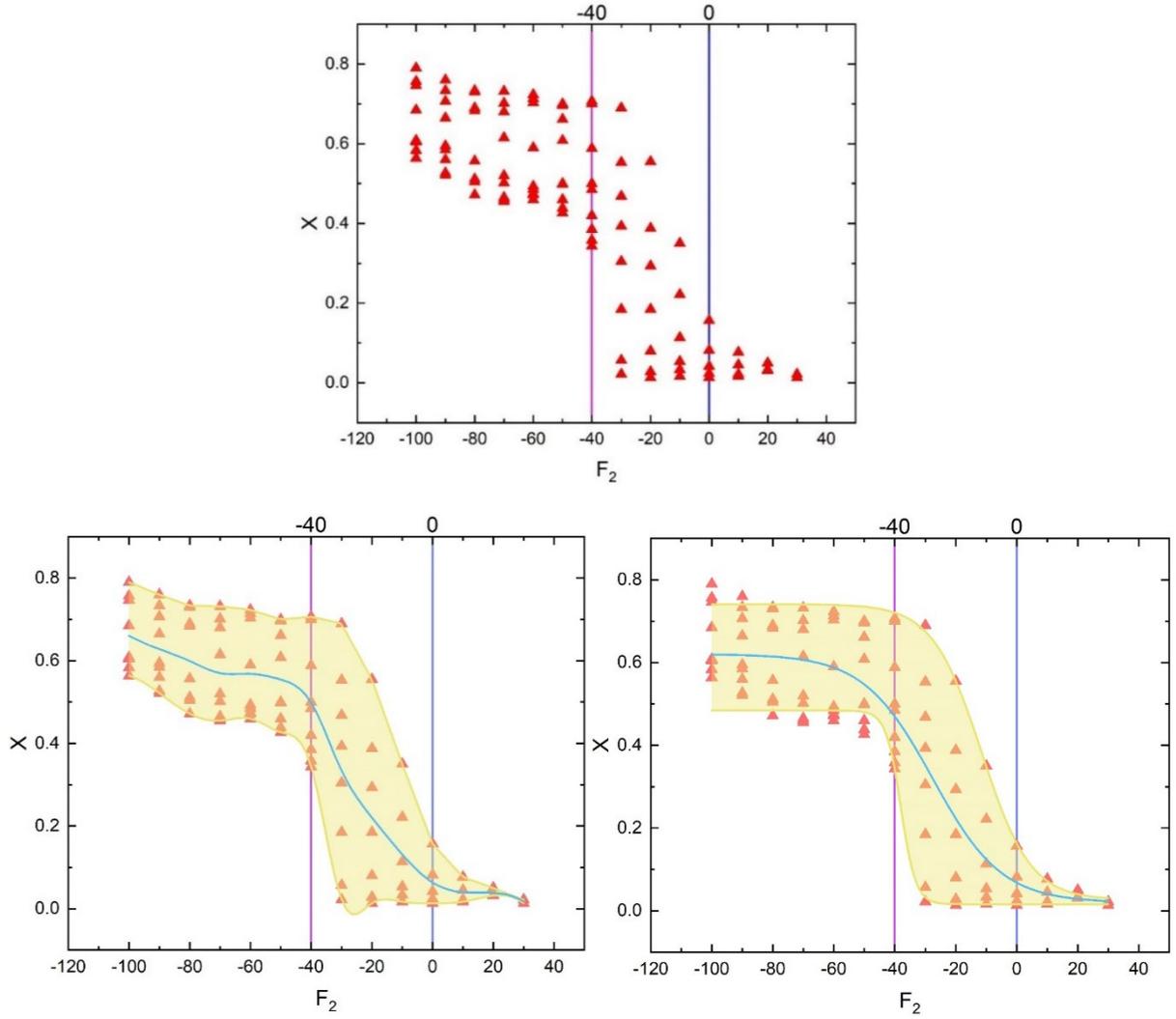

FIG. 4. The degree of adsorption on solid surface as a function of adsorption factor $F_2 = A_{22} + A_{13} - A_{12} - A_{23}$. It is seen that when $F_2 < -40$, the degree of adsorption is diffusely distributed at a high level with an overall monotonically decreasing relationship to $F_2$, while when $F_2 > 0$, $X$ sharply reduced and approaches to 0. In the transition zone between -40 and 0, $X$ fluctuates in a wide range from 0.013 to 0.69.

From FIG. 4, it is shown that as $F_2$ increases from -100 to 30, the particle distribution varies from the state with significant adsorption on the solid surface to the state without considerable adsorption, which agrees with our theory. As predicted before, the transition point $K_c$ within the interval [-40,0] is negative. In our analysis, the dilute dispersion condition is employed and it is assumed that no neighboring type 1 particles exist. However, FIG. 3(b) shows that on the solid surfaces, type 1 particles aggregate and stack with each other, which does not satisfy our assumptions and may be one of the reasons for inaccurate predictions within the transition zone. Generally, the above results demonstrate the validity of our theory in controlling the particle adsorption on solid surfaces, indicating that the additional force fields introduced to construct the solid model did not influence the thermodynamics of adsorption processes. In the next section, the adsorption process at liquid-liquid interfaces will be investigated.

## IV.C Extraction and Adsorption at Liquid-Liquid Interfaces

In contrast to the solid-liquid interface in Sec. IV.B, here a ternary mDPD system with a liquid-liquid interface was constructed, consisting of 600 type 1 particles as dispersed phase, 6401 type 2 particles and 6399 type 3 particles as two immiscible liquid phases. The simulation box was of size $33.8 \times 2.0 \times 33.8$ in reduced units and periodic boundary conditions are applied in three dimensions. In this section, the system was successively employed to investigate extraction and adsorption processes.

To be specific, simulations on 130 different models with $F_1 > 0$ and $F_3$ ranging from -55 to 45 were first conducted to reveal the relation between the degree of extraction and $F_3$, . Then, 85 models with $F_2$ ranging from -40 to 50 and $F_1 > 0$, $F_3 > 10$ to prevent agglomeration and extraction were constructed to investigate the adsorption at liquid-liquid interface. As in Sec. IV.B, all type 1 particles in the initial state were uniformly dispersed in phase 2 as shown in FIG. 5(a), after which the system evolved to the equilibrium state.

The degree of adsorption at liquid-liquid interface is quantified in the same way as in the last section, by calculating the proportion of type 1 particles with a distance less than $r_c$ from the liquid-liquid interface

$$X = \frac{N^{L}}{N^{(1)}}, \quad (27)$$

where $N^L$ is the number of type 1 particles adsorbed at the liquid-liquid interface and $N^L$ is the total number of type 1 particles. To quantify the degree of extraction, the proportion of type 1 particles immersed in phase 3 is calculated

$$I = \frac{N^{E}}{N^{(1)}}, \quad (28)$$

where $N^E$ is the number of type 1 particles extracted into phase 3.

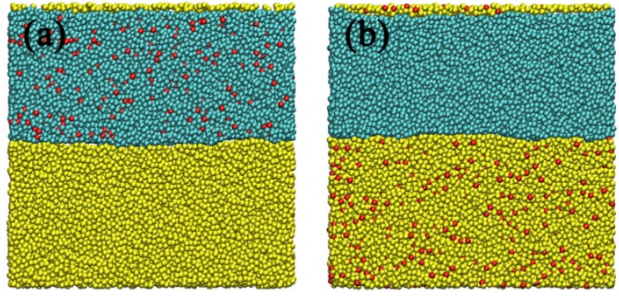

FIG. 5. The particle distribution: (a) without extraction; (b) with extraction.

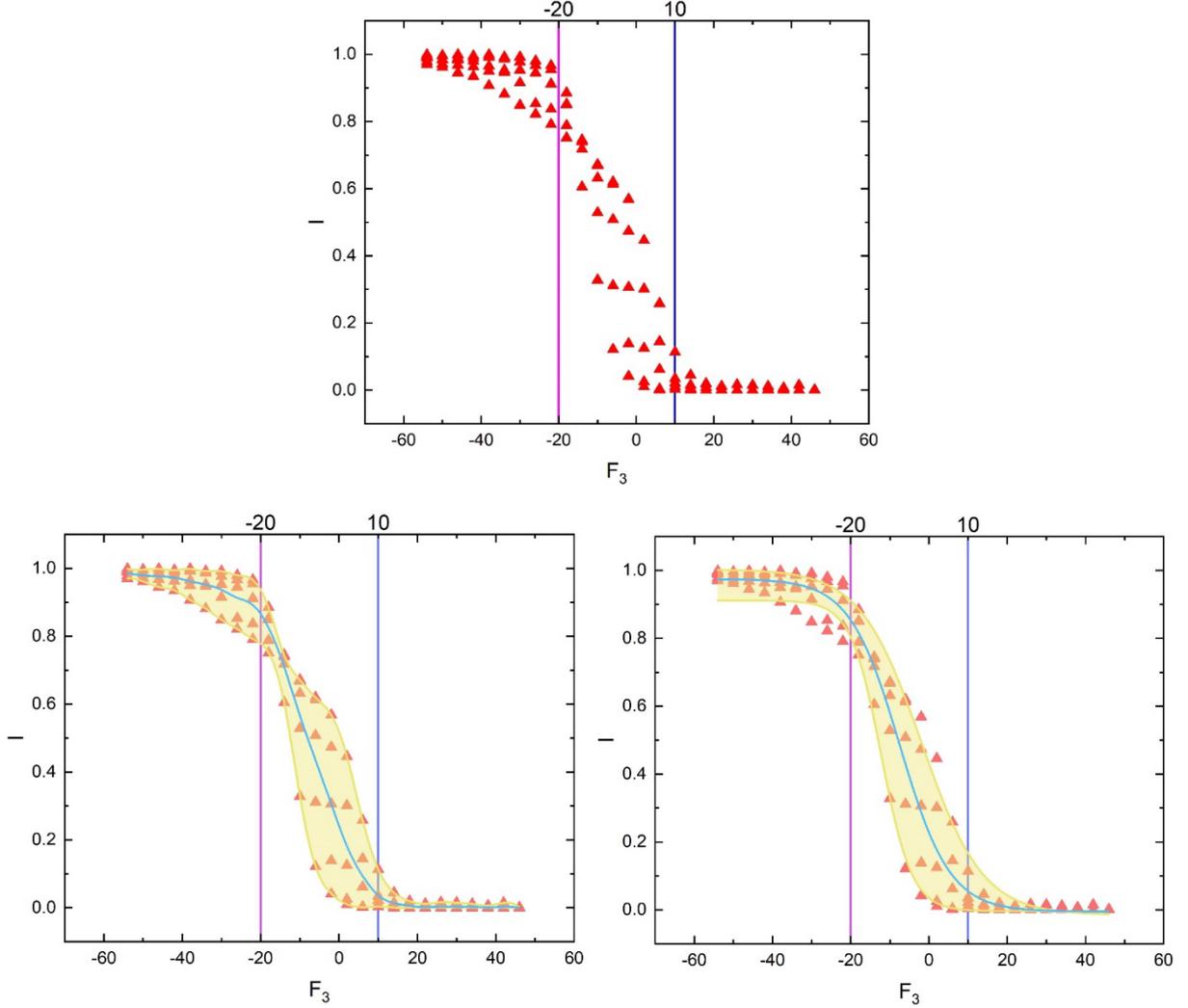

FIG. 6. The degree of extraction as a function of the extraction factor $F_3 = A_{22} + 2A_{13} - A_{33} - 2A_{12}$. It is seen that $I$ stays at a high level when $F_3 < -20$ and converges to 0 when $F_3 > 10$, while sharply decreasing with $F_3$ in the interval [-20,10].

The pattern presented in FIG. 6 is similar to those in previous sections, i.e., This pattern indicates the phase transfer of type 1 particles from phase 2 to phase 3 as $F_3$ decreases, as expected based on our model. Due to the close volume of phase 2 and phase 3, the sign of $K_c$ is indeterminate, which is compatible with our theory.

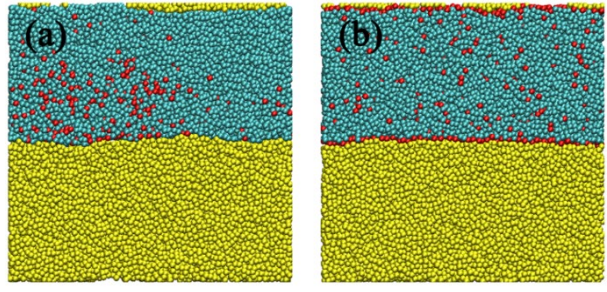

FIG. 7. The particle distribution: (a) without adsorption at liquid-liquid interface; (b) with adsorption at liquid-liquid interface.

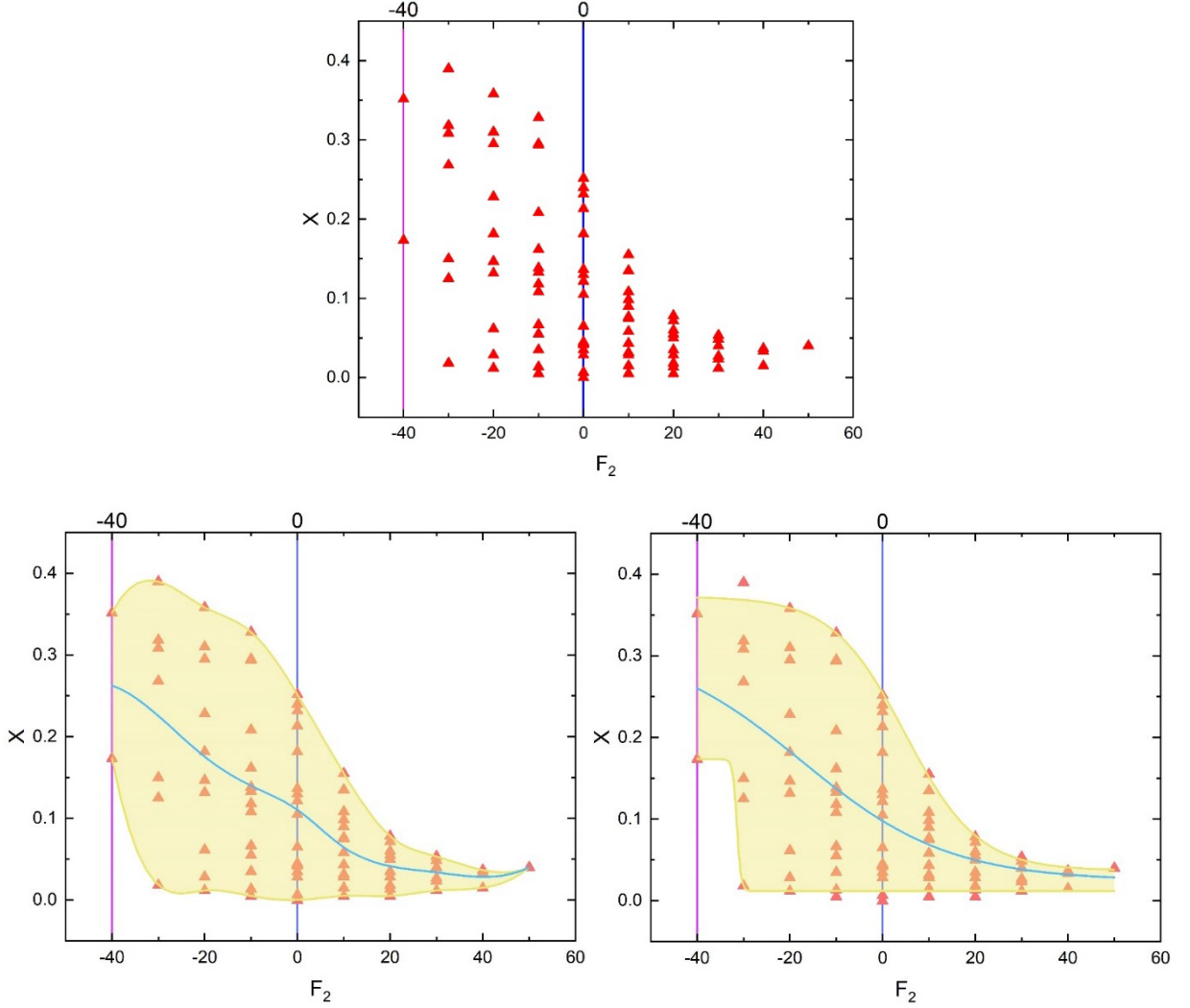

FIG. 8. The degree of adsorption at liquid-liquid interface as a function of the adsorption factor $F_2 = A_{22} + A_{13} - A_{12} - A_{23}$.

When $F_2 > 0$, $X < 0.2$ and decreases to 0, in accordance with the pattern in solid surface adsorption, indicating no significant adsorption at liquid-liquid interface when $F_2 > 0$. When $-40 < F_2 < 0$, $X$ is scattered in a wide range from 0 to 0.39, consistent with the distribution in the transition zone in Sec. IV.B.

It is indicated from FIG. 8 that $K_c < 0$, as predicted by our theory. Whereas, here only models with $F_2 > -40$ were constructed and the distribution transition of the system was not fully presented. This limitation is caused by the fact that under the premise of no extraction, if we reduce $F_2$ to below -40, $A_{23}$ will be increased to greater than 10 due to the coupling between $F_2$ and $F_3$, resulting in a significant decrease in the particle number density near the liquid-liquid interface and contradicting the assumption of an approximately uniform system density in our theoretical model. Hence, when investigating the adsorption at liquid-liquid interface, the range of $F_2$ is limited, in contrast to solid surface adsorption, where no restriction of the particle phase transfer is needed.

## V Conclusions

In this work, we extended Meyer's lattice model to describe three-component dispersed multiphase systems and proposed a general approximate form for the Helmholtz free energy of the system. Criterions based on the determinants $\Delta_i = F_i - K_{c,i}$ are accordingly proposed to predict the aforementioned behaviors based on a Helmholtz free energy formulation, with the energy factors $F_i$'s describing the internal energy effect and the entropy factors $K_{c,i}$'s depending on the local lattice configuration. To verify our prediction, a binary mDPD system, a ternary mDPD system with liquid-solid interface, and a ternary mDPD system with liquid-liquid interface were constructed to simulate the phase transition phenomena in various many-body dissipative systems, including agglomeration, adsorption, and extraction, quantified by the relative standard deviation of the number density of dispersed particles, the proportion of dispersed particles adsorbed at the interface, and the proportion of dispersed particles extracted by another continuous phase, respectively. The simulation results show the notable dependence of the dispersed phase distribution transition on the quasi-order parameters.

The good agreement between our theoretical predictions and the mDPD simulation results shows great potential of our model to further applications in dispersed multiphase systems. For example, the entropy effect is qualitatively included in the phase transition criterions, which are expected to be extended to study the structure effect induced by the temperature, phase density, and dispersed particle shape on the phase transition phenomena in those complex systems. Considering the feature of locality in our model due to the assumption of a uniform number density across the system, modeling of systems with a significant number density fluctuation or great particle size difference is fascinating and challenging. Besides, as a widely used simulation method, molecular dynamics (MD) is also widely-used in modeling multi-component particle systems similar to mDPD. Thus, applying our model to MD systems will potentially contribute to simulation studies in biology, chemicals, materials, and so on [35].

**Funding.** This work is financially supported by the NSF grant of China (No. U1837602), the National Key R&D Program of China (No. 2019YFA0708704) and the Tsinghua University Initiative Scientific Research Program.

**Declaration of interests.** The authors declare that they have no known competing financial interests or personal relationships that could have appeared to influence the work reported in this paper.

# Appendix: Detailed Calculation on the Explicit Form of the Criterion in Table 1 and 2

In this appendix, the detailed calculations on the explicit forms of the criterion of the agglomeration, adsorption, and extraction in Table 1 and 2 in Sec II.B-C, are presented successively.

## A. Table 1: Agglomeration

The agglomeration process can be investigated simply in binary systems. Consider the extended lattice model of a mixture which comprises a dispersed phase of $N^{(1)}$ type 1 particles and a continuous phase of $N^{(2)}$ type 2 particles, where $N^{(1)} \ll N^{(2)}$ is satisfied. The individual components in $\{N_1^{ij}\}$ are not independent of each other but constrained by certain geometric relations, so the independent variables should be determined.

Similar to the derivation in previous work [23], we obtain

$$N_1^{11} + N_1^{22} + N_1^{12} = \frac{c_1\left(N^{(1)} + N^{(2)}\right)}{2} \tag{29}$$

since the total number of pairs with distance $r_1$ is $c_1\left(N^{(1)} + N^{(2)}\right)/2$ and the components of each pair must be either 1-1, 2-2 and 1-2. It's worth noting that the boundary effects are neglected because $N^{(1)} + N^{(2)}$ is large. Consider each particle pair with distance $r_1$ has two "ends" belonging to the two involved sites. The total number of the "ends" that pertain to the $N^{(1)}$ sites of type 1 is $c_1 N^{(1)}$. Alternatively, this number can also be calculated by summing the contributions from 1-1 type pairs and 1-2 type pairs, namely

$$c_1 N^{(1)} = 2N_1^{11} + N_1^{12}. \tag{30}$$

Therefore,

$$N_1^{11} = \frac{\left(c_1 N_1 - N_1^{12}\right)}{2}, \quad N_1^{22} = \frac{\left(c_1 N_2 - N_1^{12}\right)}{2}. \tag{31}$$

Here, $N_1^{11}$ and $N_1^{22}$ are expressed as functions of the only independent variable $N_1^{12}$, while $N_1^{11'}$ and $N_1^{22'}$ depending on $N_1^{12'}$ in the same way. Based on Eq. (6), the Helmholtz free energy difference is rewritten as Eq. (7).

Consider that the configuration A corresponds to the uniform distribution of type 1 particles in the continuous phase, while the configuration B representing the state of agglomeration. It can be inferred that $\Delta N_1^{12}$ is positive because more contact is formed between sites of type 1 and type 2 in configuration A. Meanwhile, the higher degree of mixing in configuration A causes a greater entropy, indicating that $\Delta S$ is also positive. Then after defining the transition point $K_{c,1} = -\dfrac{T\Delta S}{\Delta N_1^{12}/2}$, which is a negative constant, it can be predicted from Eq. (7) that when $u_\alpha^{11} + u_\alpha^{22} - 2u_\alpha^{12} > K_{c,1}$, no significant agglomeration will be observed, while $u_\alpha^{11} + u_\alpha^{22} - 2u_\alpha^{12} < K_{c,1}$, considerable agglomeration will occur.

## B. Table 1: Adsorption

The adsorption process involves adsorbent particles and the interface of two immiscible phases. Consider the extended lattice model of a mixture which comprises a dispersed phase of $N^{(1)}$ type 1 particles and two continuous phases composed of $N^{(2)}$ type 2 particles and $N^{(3)}$ type 3 particles respectively, where $N^{(1)} \ll N^{(2)} \sim N^{(3)}$ is satisfied. To obtain the independent variables in this ternary system, we extend our discussion about the binary system in Sec. A.1 and get

$$N_1^{11} + N_1^{12} + N_1^{13} + N_1^{22} + N_1^{23} + N_1^{33} = \frac{c_1\left(N^{(1)} + N^{(2)} + N^{(3)}\right)}{2}, \qquad (32)$$

since the total number of pairs with distance $r_1$ is $c_1\left(N^{(1)} + N^{(2)} + N^{(3)}\right)/2$ and the components of each pair must be either 1-1, 1-2, 1-3, 2-2, 2-3 and 3-3. Similarly, the boundary effects are neglected owing to $N^{(1)} + N^{(2)} + N^{(3)}$ is large. In this case, the relations between $N^{(i)}$ and $N_1^{ij}$ are written as

$$\begin{cases} c_1 N^{(1)} = 2N_1^{11} + N_1^{12} + N_1^{13}, \\ c_1 N^{(2)} = 2N_1^{22} + N_1^{12} + N_1^{23}, \\ c_1 N^{(3)} = 2N_1^{33} + N_1^{13} + N_1^{23}, \end{cases} \qquad (33)$$

Therefore,

$$\begin{cases} N_1^{11} = \dfrac{\left(c_1 N^{(1)} - N_1^{12} - N_1^{13}\right)}{2}, \\ N_1^{22} = \dfrac{\left(c_1 N^{(2)} - N_1^{12} - N_1^{23}\right)}{2}, \\ N_1^{33} = \dfrac{\left(c_1 N^{(3)} - N_1^{13} - N_1^{23}\right)}{2}. \end{cases} \quad (34)$$

Here, $N_1^{11}$, $N_1^{22}$, and $N_1^{33}$ are expressed as functions of three independent variables $N_1^{12}$, $N_1^{23}$, and $N_1^{13}$, while $N_1^{11'}$, $N_1^{22'}$, and $N_1^{33'}$ depending on $N_1^{12'}$, $N_1^{23'}$, and $N_1^{13'}$ in the same way. Based on the Eq. (6), the Helmholtz free energy difference is rewritten as Eq. (8).

We specify that the configuration A corresponds to the uniform distribution of type 1 particles in continuous phase 2, while configuration B representing the state when all type 1 particles are adsorbed on the interface between phase 2 and phase 3. Consider the adsorption process of a single particle, which can be described by exchanging a type 1 particle within the phase 2 with a type 2 particle at the 2-3 interface, during which the variations of each $N_1^{ij}$ are

$$\Delta_0 N_1^{12} = -\Delta_0 N_1^{13} = \Delta_0 N_1^{23} = 1. \quad (35)$$

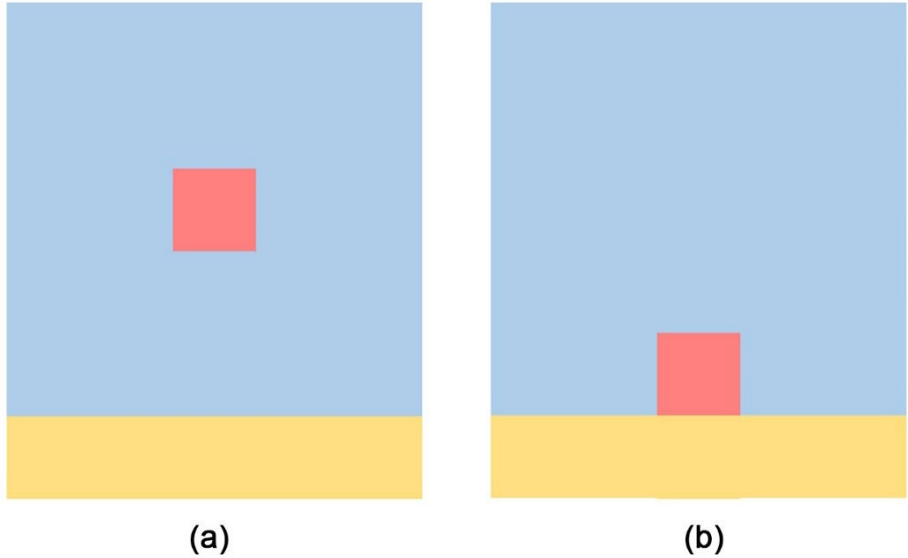

FIG. 9. Adsorption process of a single particle: (a) before adsorption; (b) after adsorption.

By repeating the above process $N^{(1)}$ times, all type 1 particles are adsorbed and we obtain $\Delta N_1^{ij}$ between configuration A and B, namely $\Delta N_1^{ij} = N^{(1)} \Delta_0 N_1^{ij}$. Here the dilute dispersion condition, $N^{(1)} \ll N^{(2)} \sim N^{(3)}$, is employed, so there are no neighboring type 1 particles. And from Eq. (8), we obtain

$$\Delta F = -(u_1^{22} + u_1^{13} - u_1^{12} - u_1^{23})N^{(1)} - T\Delta S. \quad (36)$$

It can be inferred that $\Delta S$ is positive because the lower degrees of freedom of type 1 particles when confined at the interface in configuration B, leading to a lower entropy. Hence in this case the transition point $K_{c,2} = -\dfrac{T\Delta S}{N^{(1)}}$ is also a negative constant, and it can be predicted from Eq. (36) that when $u_1^{22} + u_1^{13} - u_1^{12} - u_1^{23} > K_{c,2}$, no noticeable adsorption will be observed, while $u_1^{22} + u_1^{13} - u_1^{12} - u_1^{23} < K_{c,2}$, massive type 1 particles will adsorb on the interface.

## C. Table 1: Extraction

Same as adsorption, extraction process involves a ternary system with one dispersed phase and two continuous phases. The same extended model in the last subsection is reused and hence Eq. (8) is also applicable in this section.

Here we specify that the configuration A corresponds to the uniform distribution of type 1 particles in continuous phase 2, while configuration B representing the state when all type 1 particles are dispersed in phase 3. Consider the extraction process of a single particle, which can be described by moving a type 1 particle within the phase 2 into the interior of the phase 3, during which the variations of each $N_1^{ij}$ are

$$\Delta_0 N_1^{12} = -\Delta_0 N_1^{13} = 6, \Delta_0 N_1^{23} = 0. \tag{37}$$

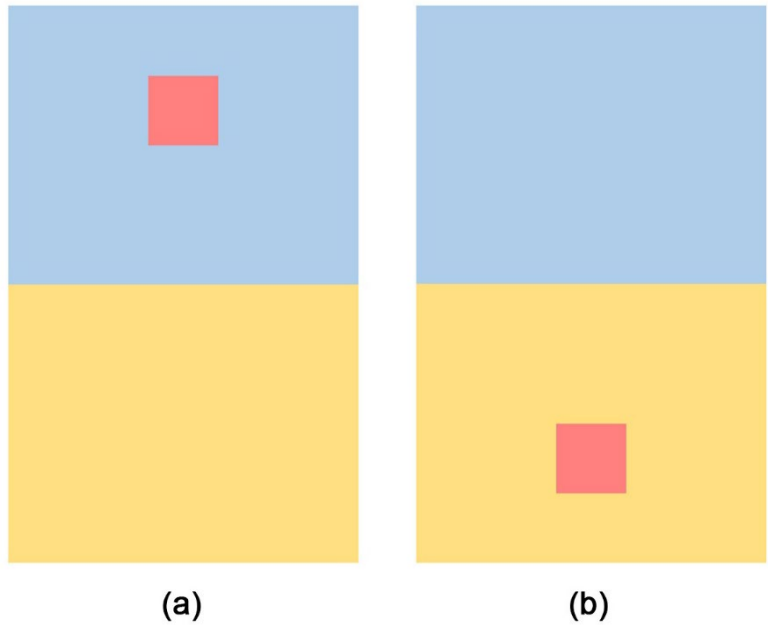

FIG. 10. Extraction process of a single particle: (a) before extraction; (b) after extraction.

By repeating the above process $N^{(1)}$ times, all type 1 particles are extracted into phase 3 and we obtain $\Delta N_1^{ij}$ between configuration A and B, namely $\Delta N_1^{ij} = N^{(1)} \Delta_0 N_1^{ij}$. Again, the assumption that no neighboring type 1

pair exists is made. Based on Eq. (8), we have

$$\Delta F = -3(u_1^{22} + 2u_1^{13} - u_1^{33} - 2u_1^{12})N^{(1)} - T\Delta S. \qquad (38)$$

The sign of $\Delta S$ is determined by the ratio of $N^{(2)}$ and $N^{(3)}$, that is, when $\dfrac{N^{(2)}}{N^{(3)}} > 1 \left( \dfrac{N^{(2)}}{N^{(3)}} < 1 \right)$, $\Delta S > 0 \, (\Delta S < 0)$. Hence the transition point $K_{c,3} = -\dfrac{T\Delta S}{3N^{(1)}}$ is a constant with its sign depending on $\dfrac{N^{(2)}}{N^{(3)}}$.

Then it can be predicted from Eq. (38) that when $u_1^{22} + 2u_1^{13} - u_1^{33} - 2u_1^{12} > K_{c,3}$, no noticeable extraction will be observed, while $u_1^{22} + 2u_1^{13} - u_1^{33} - 2u_1^{12} < K_{c,3}$, massive type 1 particles will be extracted into phase 3.

### D. Table 2: Adsorption and extraction

Consider a ternary mixture with a liquid-liquid interface. For adsorption process, consider that a type 1 particle within the phase 2 far from the interface is exchanged with a type 2 particle at the 2-3 interface, during which the variations of each $N_\alpha^{ij}$ satisfy the following relationship

$$\Delta_0 N_\alpha^{12} = -\Delta_0 N_\alpha^{13} = \Delta_0 N_\alpha^{23}. \qquad (39)$$

Hence,

$$\Delta N_\alpha^{12} = -\Delta N_\alpha^{13} = \Delta N_\alpha^{23}. \qquad (40)$$

Combine Eq. (14) and Eq. (40), the free energy change for the adsorption process is obtained

$$\Delta F = -(A^{22} + A^{13} - A^{12} - A^{23}) \sum_{\alpha=1}^{\infty} \left( w(r_\alpha) \Delta N_\alpha^{12} \right) - T\Delta S, \qquad (41)$$

Eq. (41) indicates that the distribution transition for adsorption is controlled by the potential coefficient combination $F_2 = A^{22} + A^{13} - A^{12} - A^{23}$. And the transition point becomes $K_{c,2} = -\dfrac{T\Delta S}{\sum_{\alpha=1}^{\infty} \left( w(r_\alpha) \Delta N_\alpha^{12} \right)}$.

In the same manner, for extraction process the variations of each $N_\alpha^{ij}$ satisfy the following relationship

$$\Delta N_1^{12} = -\Delta N_1^{13}, \, \Delta N_1^{23} = 0. \qquad (42)$$

And the free energy difference is expressed as:

$$\Delta F = -(A^{22} + 2A^{13} - A^{33} - 2A^{12}) \sum_{\alpha=1}^{\infty} \left( w(r_\alpha) \dfrac{\Delta N_\alpha^{12}}{2} \right) - T\Delta S. \qquad (43)$$

Here the potential coefficient combination $F_3 = A^{22} + 2A^{13} - A^{33} - 2A^{12}$ is the critical factor that determines whether the extraction happens and the transition point is changed to $K_{c,3} = -\dfrac{T\Delta S}{\sum_{\alpha=1}^{\infty}\left(w(r_\alpha)\dfrac{\Delta N_\alpha^{12}}{2}\right)}$.